\documentclass[%
 reprint,
superscriptaddress,
 amsmath,amssymb,
 aps,
]{revtex4-1}

\usepackage[colorlinks]{hyperref}

\usepackage{mathrsfs}
\usepackage{graphicx}% Include figure files
\usepackage{dcolumn}% Align table columns on decimal point
\usepackage{bm}% bold math

\newcommand{\bea}{\begin{eqnarray}}
\newcommand{\eea}{\end{eqnarray}}
\newcommand{\be}{\begin{equation}}
\newcommand{\ee}{\end{equation}}
\newcommand{\rti}{\tilde{r}}
\newcommand{\tti}{\tilde{t}}

\begin{document}

\title{A toy model for a baby universe inside a black hole}

\author{Hrishikesh~Chakrabarty}
\email{chrishikesh17@fudan.edu.cn}
\affiliation{Center for Field Theory and Particle Physics and Department of Physics, Fudan University, 200438 Shanghai, China}

\author{Ahmadjon Abdujabbarov}
\email{ahmadjon@astrin.uz}
\affiliation{Shanghai Astronomical Observatory, 80 Nandan Road, Shanghai 200030, P. R. China}
\affiliation{Ulugh Beg Astronomical Institute, Astronomicheskaya 33, Tashkent 100052, Uzbekistan}
\affiliation{National University of Uzbekistan, Tashkent 100174, Uzbekistan}

\author{Daniele~Malafarina}
\email{daniele.malafarina@nu.edu.kz}
\affiliation{Department of Physics, Nazarbayev University, 53 Kabanbay Batyr avenue, 010000 Astana, Kazakhstan}

\author{Cosimo~Bambi}
\email[Corresponding author: ]{bambi@fudan.edu.cn}
\affiliation{Center for Field Theory and Particle Physics and Department of Physics, Fudan University, 200438 Shanghai, China}

\date{\today}% It is always \today, today,

\begin{abstract}
We present a dynamical toy model for an expanding universe inside a black hole. The model is built by matching a spherically symmetric collapsing matter cloud to an expanding Friedmann-Robertson-Walker universe through a phase transition that occurs in the quantum-gravity dominated region, here modeled with semi-classical corrections at high density. The matching is performed on a space-like hyper-surface identified by the co-moving time at which quantum-gravity induced effects halt collapse. 
The purpose of the model is to suggest a possible reconciliation between the observation that black holes are well described by the classical solutions and the fact that the theoretical resolution of space-time singularities leads to a bounce for the collapsing matter.
\end{abstract}

\maketitle

%\tableofcontents

\section{Introduction}

Singularities appear to be unavoidable as the endstate of complete gravitational collapse within General Relativity (GR) under physically reasonable conditions \cite{sing,sing2}. The fact that the existence of curvature singularities leads to
a breakdown of the causal structure of the space-time with loss of predictability has led Penrose to conjecture that such pathologies must always be hidden behind horizons \cite{ccc}.

Another way to address the problem relies on the possibility that singularities must be generically resolved within a theory of quantum-gravity (QG) and therefore do not arise in realistic physical scenarios 
\cite{QG}.
This idea bears significant consequences for both cosmology \cite{hw,ash} and black hole physics (see \cite{dm} for a recent review).
Dynamical models for `quantum inspired' bounces from gravitational collapse have become of great interest in recent years as they pose interesting theoretical questions on the consequences of singularity resolution within black holes and their possible implications for astrophysics
\cite{liberati}.
The most important feature of such models is that, by removing the central singularity that forms at the end of collapse in the classical scenario, they allow for the space-time to be geodesically complete. Matter and information are not destroyed in the singularity, leaving the door open for other possibilities
(see for example \cite{Frolov}-\cite{baccetti}).
However, the nature of the bounce and the evolution of the matter fields after the bounce are at present poorly understood and several theoretical questions remain open \cite{dm}.

In particular one of the main open problems is the nature of the geometry of the space-time in the exterior of the collapsing cloud after the bounce. The recent observation of the shadow of the supermassive black hole candidate at the center of the galaxy M87 by the Event Horizon Telescope collaboration has not shown any significant departure from the prediction of GR
\cite{EHT1,EHT2}.
At present, from theoretical consideration, it is clear that the QG modifications that allow for the resolution of the singularity must affect the geometry at distances larger than the Planck scale and possibly outside the boundary of the collapsing object.
However, it is unclear whether such modifications may propagate to the exterior of the event horizon
\cite{rov}. 
It has been suggested that such modifications may extend outside the horizon but must remain confined within a short (i.e. Planck scale) distance
(see for example \cite{gid} or \cite{Carballo-Rubio}, \cite{bonanno} for models where the effects are macroscopic). 

In the present work, we assume that no departure from the classical black hole solution exists outside the horizon and investigate one possible behavior of matter in the post-bounce era by constructing a toy model that leads to a baby universe causally disconnected from its parent universe.
The idea of the existence of universes inside black holes has been suggested by many authors.
For example in
\cite{smolin}
it was suggested that universes with black holes arise naturally through a selection process where new universes with slightly different values of fundamental constants are created inside black holes. A critical analysis of this idea was developed in
\cite{ellis}
and the implications of the idea for longstanding problems in cosmology were analyzed in
\cite{bran}.
In
\cite{FMM1,FMM2}
the authors matched the interior of a Schwarzschild black hole at the endpoint of collapse with a DeSitter universe.
Other models with new universes being created in regions of space-time causally disconnected from the parent universe were considered for example in
\cite{HL}-\cite{dom2}.

In our model, we follow a similar approach to that of \cite{FMM1} with the additional ingredient of the bounce. Therefore, we match a collapsing matter cloud to an expanding Friedmann-Robertson-Walker (FRW) universe at a space-like hyper-surface defined by the time at which collapse halts and the bounce occurs. 
This addition of the bounce mechanism naturally allows for an expanding universe to develop inside the black hole, thus supporting Smolin's idea of universes with black hole being favoured via a selection mechanism.
The geometry outside the horizon is assumed to be described by Schwarzschild at all times.  
In this way, the observers at infinity will only detect the formation of a classical black hole without any evidence of the bounce. 

As noted by several authors, the price to pay for allowing the collapsing matter to bounce, thus avoiding the formation of the singularity, is the necessity for the presence of a finite region of space-time where classical GR does not hold. This is due to the fact that the light-cone structure of the space-time must change in the transition of the geometry from collapsing to expanding, and thus a finite size region interpolating between the two geometries must exist (see for example \cite{rov}).
In our case, such region, located in the vicinity of the space-like matching surface, remains confined inside the black hole horizon.

It must be noted that the model does not describe the mechanism of the phase transition at the bounce, since that would require a quantum gravitational theory of matter. Therefore the model does not predict what matter content would be allowed in the expanding phase. However, we show that if such a phase transition happens, then the bounce mechanism provides a natural framework for the extension of the geometry in the expanding universe.

The paper is organized as follows:
In section \ref{coll} we construct the model for collapse matched to an expanding FRW metric. Sections \ref{exterior} and \ref{expand} are devoted to the discussion of the matching of the exterior vacuum space-time with the collapsing interior and with the expanding universe, respectively with particular attention to the behavior of the apparent horizon. Finally in section \ref{discuss} we outline some properties and open issues of the model presented.

Throughout the paper, we make use of natural units setting $G=c=1$.

\section{Dust collapse with bounce matched to FRW}\label{coll}

We shall begin by considering the collapsing interior space-time (we shall call it region I) described by homogeneous dust, i.e. the Oppenheimer-Snyder-Datt (OSD) model 
\cite{OS,datt}. We shall label the coordinates as $\{t_I, r_I, \theta, \phi \}$. The equation of motion for the classical collapsing homogeneous dust cloud is~\cite{collapse}
\begin{equation}\label{eq1}
    \dot{a}_I=-\sqrt{\frac{m_I}{a_I}-k_I},
\end{equation}
where the dot represents derivative with respect to the co-moving time $t_I$ and $ m_I $ and $k_I$ are constants related to the initial density and initial velocity of the collapsing dust cloud. 
The scale factor $a_I(t_I)$ is the only degree of freedom of the system and describes the rate of collapse of each shell of the dust cloud.
In order to turn collapse into a bounce, one must require either a violation of the energy conditions or failure of Einstein's equations. It can be argued that in some cases the modifications to GR at high densities can be treated as an effective energy-momentum tensor thus reducing the problem to that of classical Einstein's equations with an effective matter-source describing the modifications to GR \cite{barcelo2}. 
Then the minimal requirement in order to obtain a bounce leading to a new universe is that the matter source (effective or not) violates the strong energy condition
\cite{visser2}.
In \cite{BMM} it was shown how the OSD model can be modified in order to halt collapse at a finite radius and obtain a matter bounce if an effective energy density is introduced in the form
\be 
\rho^{\rm eff}=\rho-\frac{\rho^2}{\rho_{\rm cr}},
\ee 
where $\rho_{\rm cr}$ is a critical density that signals a regime where the classical description fails.
{By replacing the physical density with the effective density one obtains a model described in terms of effective quantities. These are the effective density $\rho^{\rm eff}(t)$, pressure $p^{\rm eff}(t)$ and mass $M^{\rm eff}(t)$. Then one obtains a model of collapse in which the equations for the effective quantities are formally identical to the usual Einstein's equations for a perfect fluid. However, due to the non-physical nature of the effective corrections, the energy conditions can be violated for the effective quantities, thus allowing for collapse to halt (see \cite{BMM} for details).
}
The effective equation of motion then becomes
\begin{equation}\label{adot}
    \dot{a}_I=-\sqrt{\frac{m_I}{a_I}\left(1-\frac{\rho_I}{\rho_{\rm cr}^I}\right)-k_I}.
\end{equation}
In the simplest case of marginally bound collapse, i.e. $k_I=0$, the solution is readily obtained as
\be 
a_I(t_I)=\left[a_I(t_B)^3+\left(\sqrt{1-a_I(t_B)^3}
-\frac{3\sqrt{m_I}}{2}t_I\right)^2\right]^{1/3}.
\ee 
In this scenario the collapse halts at a size $a_I(t_B)=(3m_I/\rho_{\rm cr})^{1/3}$ when $\dot{a}_I=0$ and $\rho_I=\rho_{\rm cr}$. This occurs at the time
\be
t_I=t_B=\frac{2\sqrt{1-(3m_I/\rho_{\rm cr})^{1/3}}}{3\sqrt{m_I}}.
\ee
In the following, we shall consider the more general and more realistic case of bound collapse (i.e. $k_i>0$) where the `star' initiates collapse with zero velocity at a finite radius. 
This is the case also classically, as can be seen from equation \eqref{eq1} for which the scale factor must satisfy $a\leq m/k$. Classically $a=m/k$ is the only root of $\dot{a}$=0 and must define the initial configuration. On the other hand, in the quantum-inspired model we see from equation \eqref{adot} that $\dot{a}^2=0$ is a quartic function of $a$ that has two positive real solutions 
for 
$m>(16k/3)\sqrt{k/\rho_{\rm cr}}$
(for the sake of clarity in this paragraph we have omitted the subscript). Since we are considering collapse of massive objects (such as stars) which is halted in the quantum-gravity regime, which can be assumed to have $\rho_{\rm cr}$ of the order of the Planck density, it is easy to see that for any reasonable choice of $m$, $k$ and $\rho_{\rm cr}$ equation \eqref{adot} will have two positive roots.
Then the study of the dynamical system $f(a)=a^4-ma^3/k+3m^2/(k\rho_{\rm cr})$ obtained from equation \eqref{adot} shows that motion is allowed within the two positive solutions of $f(a)=0$ given by the initial time and the time of the bounce. The behavior of collapse is similar to the marginally bound case and the cloud halts and bounces back at a finite co-moving time $t_I=t_B$. In this case, the maximum density achieved at the time of the bounce $t_B$ is 
\be 
\rho_B=\left(1-\frac{ka(t_B)}{m}\right)\rho_{\rm cr}<\rho_{\rm cr}.
\ee

We shall now treat the hyper-surface $\Sigma$ given by $t_I=t_B$ as a phase transition at which the collapsing matter turns into an expanding FRW universe. We shall call the expanding phase region II and label the coordinates as $\{t_{II}, r_{II}, \theta, \phi \}$. 
The metric on both sides of the phase transition can be written as
\begin{equation}\label{metric-in}
    ds_{i}^{2}=-dt_{i}^{2}+\frac{a_{i}(t_{i})^{2}}{1-k_{i}r_{i}^{2}}dr_{i}^{2}+(r_{i}a_{i})^{2}d\Omega^{2},
\end{equation}
with $i=I, II$ and the matching across the hyper-surface can be done following the Darmois-Israel formalism (see \cite{israel}-\cite{matching3}).
The boundary hyper-surface can be given in parametric form on both sides as
\begin{equation}
  t_{i}-t_{B}=0.
\end{equation}
The induced metric on the hyper-surface, as seen from the $i$-th region is given by 
\begin{equation}
    h_{ab}^{i}=g_{\alpha\beta}^{i}e^{\alpha}_{a}e^{\beta}_{b},
\end{equation}
where, $ e^{\alpha}_{a}=\partial x^{\alpha}/\partial y^{a} $ are the tangent vectors on the hyper-surface $\Sigma$ which has coordinates $ y^a=(r,\theta,\phi) $ and Latin letters cover the three spatial coordinates.
The non-vanishing components of the induced metric are 
\be 
h_{rr}^{i}=\frac{a_{i}(t_{i})^{2}}{1-k_{i}r_{i}^{2}}, \; \; h_{\theta\theta}^{i}=\frac{h_{\phi\phi}^{i}}{\sin^2\theta}=(a_{i}r_{i})^{2},
\ee 
and the three dimensional line element on $\Sigma$ can be written in either coordinate systems as
\begin{equation}
 ds^{2}_{\Sigma}=\frac{a_{i}(t_{B})^{2}}{1-k_{i}r_{i}^{2}}dr_{i}^{2}+a_{i}(t_B)^2r_{i}^{2}d\Omega^{2}.
\end{equation}

Now, from the matching conditions $ h_{ab}^{I}=h_{ab}^{II} $, we get
\bea \label{1st}
   r_{II}&=&\frac{a_{I}(t_{B})}{a_{II}(t_{B})}r_I, \\ \label{2nd}
   a_{II}(t_{B})&=&\sqrt{\frac{k_{II}}{k_I}}a_{I}(t_{B}),
\eea
the second of which gives the initial condition for the scale factor in the expanding phase. 
To have a smooth matching across the phase transition surface one needs to ensure that the second fundamental form has the same value on both sides.
The only non-vanishing components of the extrinsic curvature as seen from either side are
\bea
K_{rr}^{i}&=&\frac{a_{i}\dot{a}_{i}}{1-k_{i}r_{i}^{2}}, \\
 K_{\theta\theta}&=&\frac{K_{\phi\phi}}{\sin^2\theta}=a_i\dot{a}_ir_i^2,
\eea
from which we see that continuous matching is possible at the surface $t_{i}=t_{B}$ since
\begin{equation}\label{ic}
     \dot{a}_i(t_{B})=0 .
\end{equation}

\begin{figure}[h]
	\begin{center}
		\includegraphics[width=0.45\textwidth]{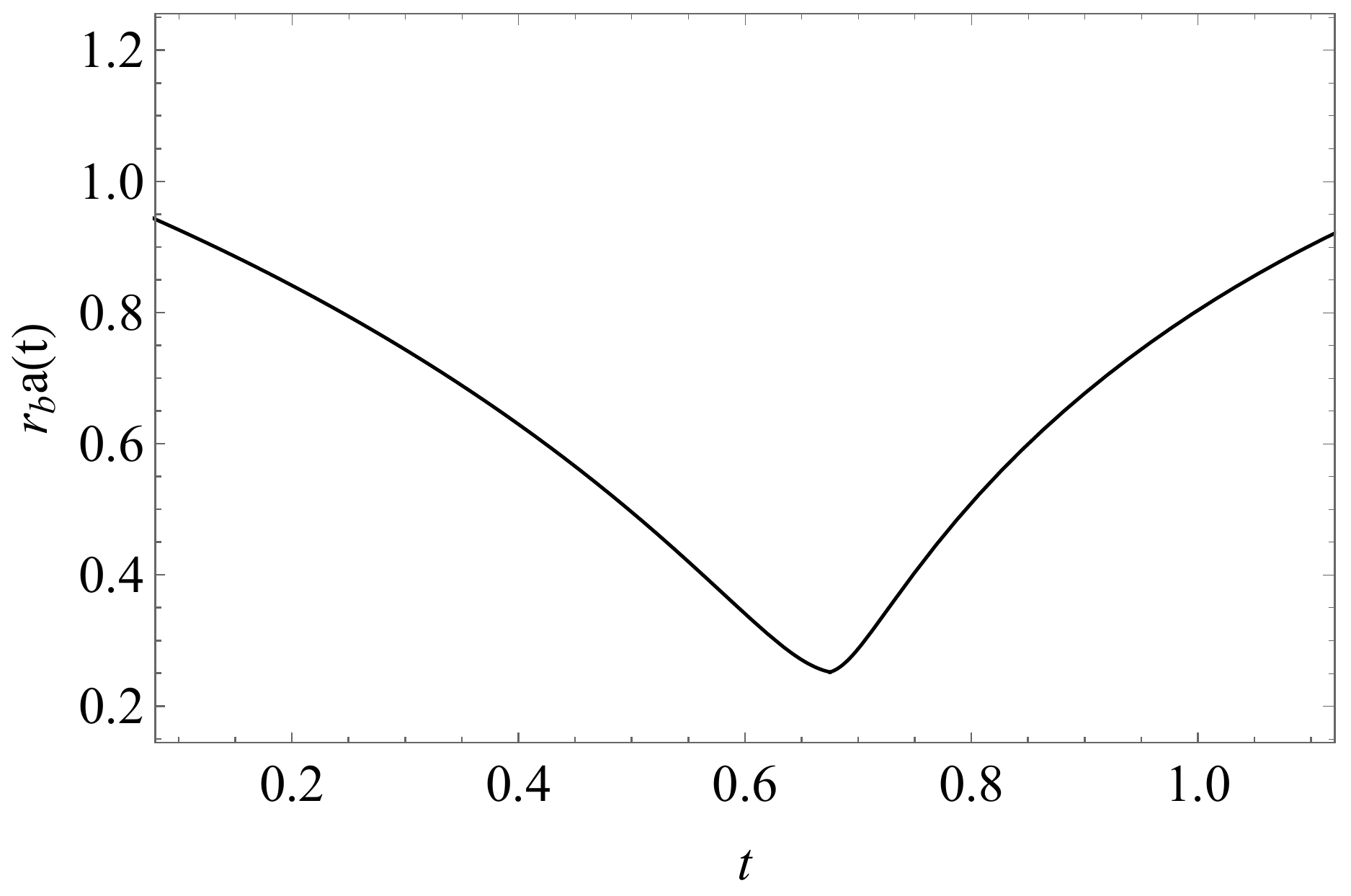}
	\end{center}
	\caption{ 
	The physical radius of the boundary $R_b(t)=r_ba(t)$ as a function of the co-moving time with $a(t)=a_I(t_I)$ for $t\leq t_B$ and $a(t)=a_{II}(t_{II})$ for $t\geq t_B$. The bounce occurs at the critical scale given by $a_I(t_B)=a_{II}(t_B)$. The values of the parameters are $ k_I = 1.5 $, $ m_I = 2 $ and $ k_{II} = 1.96 $, $ m_{II} = 3 $, $ \omega = 1/3 $ for the collapsing and the expanding part respectively.
	\label{bounce1}}
\end{figure}

In Fig.~(\ref{bounce1}), it is shown the evolution of the scale factor from the initial condition $a_I(0)=1$ (which need not correspond to the initial time at which $\dot{a}_I=0$) to the bounce and into the expanding phase. 
Considering a FRW metric with a perfect fluid energy-momentum tensor and equation of state $p_{II}=\omega\rho_{II}$ ($\omega\in[-1,1]$), we can write the equation of motion for the scale factor in the expanding region as~\cite{BMM}
\be \label{eom2}
\dot{a}_{II}=\sqrt{\frac{m_{II}}{a_{II}^{3\omega+1}}\left(1-\frac{\rho_{II}}{\rho_{\rm cr}}\right)-k_{II}},
\ee
with initial condition given by equation \eqref{2nd}. 

Then, assuming that the value of the critical density be the same on both sides, the initial condition given by equation \eqref{ic} can be satisfied at $t_{II}=t_B$ once $m_{II}$ and $\omega$ are chosen in such a way that $m_{II}/a_{II}^{3(\omega+1)}=m_I/a_I^3$ at $t_I=t_{II}=t_B$.
Note that in principle one could also assume that the critical density be different on either side of the phase transition in which case one would have to choose $m_{II}$, $\omega$ and the new $\rho_{\rm cr}$ in such a way that $\dot{a}_{II}(t_B)=0$.

\section{The exterior space-time} \label{exterior}

We now turn the attention to the exterior geometry in the collapsing phase.
In the classical picture, as the cloud collapses a black hole forms when the matter cloud passes the threshold of the horizon. In the following we assume that the modifications to the matter fields due to QG effects do not alter the exterior geometry and so for the exterior we shall consider the Schwarzschild solution with coordinates $\{T,R,\theta,\phi\}$ and mass parameter $ M_{\rm Sch} $ given by the matching of the interior mass at the boundary of the collapsing cloud. We shall call the exterior space-time region III, with line element given by
\begin{equation}\label{sch}
    ds_{III}^{2}=-FdT^{2}+F^{-1}dR^{2}+R^{2}d\Omega^{2},
\end{equation}
with
\begin{equation}
    F=1-\frac{2M_{\rm Sch}}{R}.
\end{equation}

Classically, the continuous matching of the second fundamental form across the boundary $R_{b}(T(t_I))=r_{Ib}a_I(t_I)$ implies that the total mass of the collapsing cloud is related to the Schwarzschild mass $M_{\rm Sch}$ of the exterior space-time via
\be \label{msch}
M_{\rm Sch}=\frac{1}{2}r_{Ib}^3m_I,
\ee
where the co-moving boundary $r_{Ib}$ in the region I can be chosen arbitrarily due to the absence of pressures. 
However, in our model, we have introduced semi-classical corrections for which the effective density and the effective mass of the interior space-time change because of the contribution to the geometry of QG effects that become important at high densities. Therefore the Misner-Sharp mass function in the interior is replaced by its effective counterpart which is not constant in time. In fact, it is easy to see that,
\be 
M^{\rm eff}_I(t_I)=m_I\left(1-\frac{\rho_I}{\rho_{\rm cr}}\right).
\ee 
At large radii (or early times in collapse) the corrections are negligible and in the limit of $\rho_{\rm cr}\rightarrow +\infty$, we recover the classical scenario. However, as collapse progresses the corrections become important and the effective mass in the interior is not conserved anymore so that at the boundary the effective mass is lower than the corresponding Schwarzschild mass given by equation \eqref{msch}. This may be interpreted as a sign that semi-classical effects must propagate to the exterior region.

On the other hand, here we consider the exterior space-time to be a classical black hole and assume that semi-classical corrections do not alter the exterior geometry. Therefore, in order to match the semi-classical interior with the Schwarzschild exterior, we need to allow for an effective energy-momentum tensor to be present on the boundary surface. This additional matter component must be understood as resulting from QG corrections to the interior geometry that exists only in the presence of high density matter fields.
Therefore the effective matter content on the boundary allows for the matching of a classical Schwarzschild exterior to the interior described above and must be understood not as a physical matter field but as a consequence of the semi-classical corrections in the interior.

The matching between the interior of the region I and the exterior of region III is performed across the time-like boundary surface given by $r_I-r_{Ib}=0$ in the region I and $R-R_b(T)=0$ in region III. 
Continuity of the metric is easily ensured from the conditions 
\bea
 \frac{dT}{dt_I}&=& \sqrt{\frac{r_{I}^{2}\dot{a}_{I}^{2}}{F^{2}} + \frac{1}{F}}, \\
 R_b(T(t_I))&=&r_{Ib}a_I(t_I).
\eea 
On the other hand, continuity of the second fundamental form can not be satisfied in this case. The extrinsic curvature being not same on the two sides of the collapsing boundary implies that the shell $r_I=r_{Ib}$ must carry a delta-like surface stress-energy tensor $S_{ab}$ given by
\begin{equation}\label{matching2}
     S_{ab}=-\frac{\epsilon}{8\pi}\big( [K_{ab}]-[K]h_{ab} \big),
\end{equation}
where $\epsilon=1$ for a time-like boundary, $h_{ab}$ is the induced metric on the boundary and $[K_{ab}]$ is the jump of the extrinsic curvature across the boundary. Here the notation $ [A] $ which defines the jump of $A$ across a surface $\Sigma$ is given by $[A]=A^{III}\mid_{\Sigma}-A^{I}\mid_{\Sigma}$.

\begin{figure}[h]
	\begin{center}
	\includegraphics[width=0.45\textwidth]{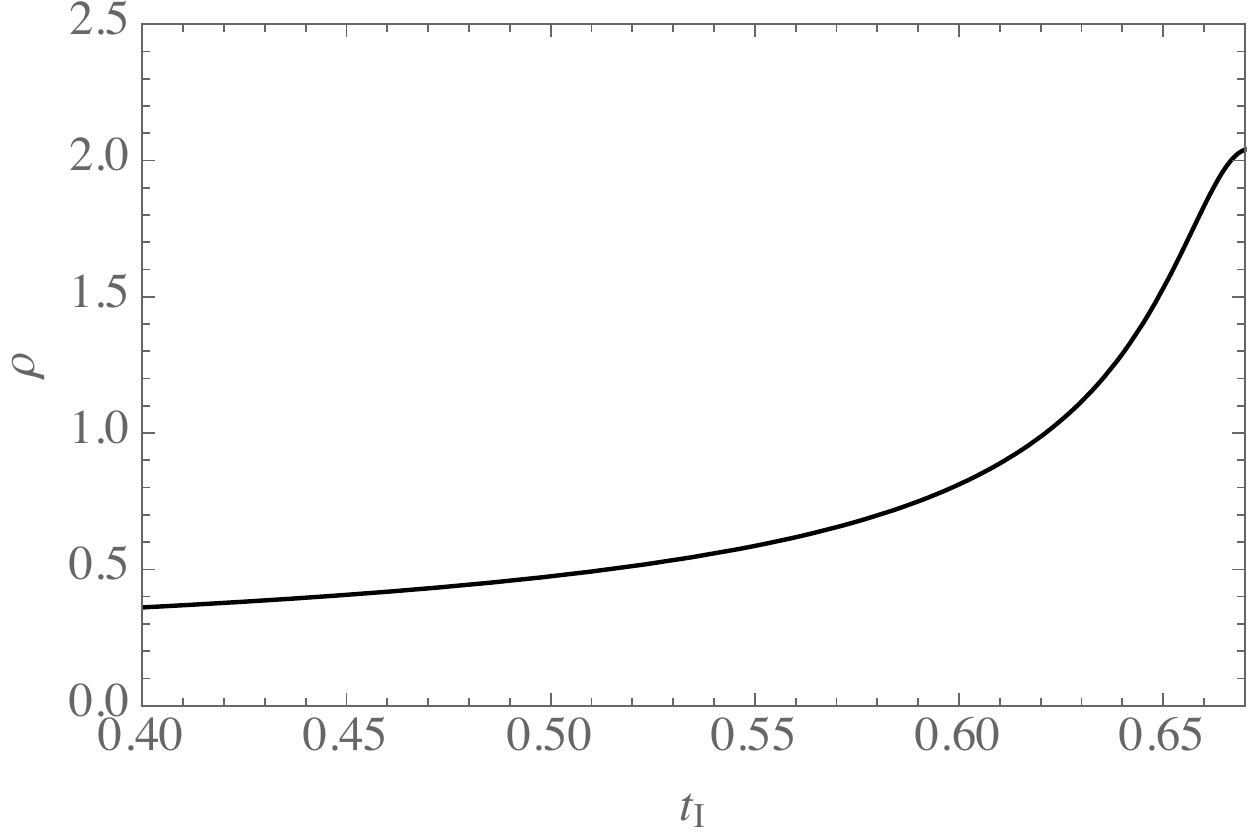}
	\end{center}
	\vspace{-0.5cm}
	\caption{Evolution of the effective density $\rho_\Sigma$ concentrated on the boundary surface between region I and region III. The shell's density increases in such a way that $M_\Sigma+r_{Ib}^3M^{\rm eff}_I/2=M_{\rm Sch}$ at all times. The numerical parameters used in these plot are $ m_I = 2 $, $ k_I = 1.5 $.
	\label{den} }
\end{figure}

For region I the extrinsic curvature is given by
\be 
K_{tt}^{I} =  0, \; \;
        K_{\theta\theta}^{I} = \frac{K^I_{\phi\phi}}{\sin^2\theta}=r_{I}a_{I}(t_I)\sqrt{1-k_Ir_I^{2}}.
\ee
For region III the extrinsic curvature is given by 
\begin{equation}
    \begin{aligned}
       & K_{tt}^{III} =\frac{-1}{F^{-1}\dot{R}_{b}-F}\bigg[ \frac{M_{\rm Sch}\dot{R}_{b}^2\dot{T}}{(2M_{\rm Sch}-R_b)R_b} \\
       & \ \ \ \ \ \ \ \  +\frac{M_{\rm Sch}(2M_{\rm Sch}-R_b)\dot{T}^{2}}{R_b^{3}}+\frac{M_{\rm Sch}\dot{R}_{b}^2}{(2M_{\rm Sch}-R_b)R_b} \bigg], \\
       & K_{\theta\theta}^{III} = \frac{K^{III}_{\phi\phi}}{\sin^2\theta}=   \frac{-(2M_{\rm Sch}-R_b)}{F^{-1}\dot{R}_{b}-F},
    \end{aligned}
\end{equation}
where dot denotes derivatives with respect to the co-moving time $t_I$ in region I.
Now from the 2nd junction condition, we obtain the effective density on the shell as
\begin{equation}
    S_{tt}=\rho_\Sigma (t_I)=-\frac{1}{8\pi}\bigg( [K_{tt}]-[K]h_{tt}^{I} \bigg),
\end{equation}
where $K=h^{ab}K_{ab}$.
The effective mass distribution on the shell is given by
\begin{equation}
    M_{\Sigma}=\rho_\Sigma \times 4\pi r_{Ib}^{2}.
\end{equation}
In figure \ref{den} we show the evolution of the effective density of the boundary surface between region I and region III.

\section{The expanding phase}\label{expand}

As the collapsing matter (I) reaches the phase transition the space-time turns into an expanding FRW universe (II). This leaves open the question of what happens to the portion of the Schwarzschild space-time between the boundary and the horizon of region III. There are several possibilities that can be constructed at this point and we will mention here three that are, in our view, the most significant: 
\begin{itemize}
    \item[(a)] Matching the Schwarzschild region III to another Schwarzschild space-time (region IV), with a different value of the mass parameter, at a surface $R=R_0=const.$ inside the horizon on both sides. 
    \item[(b)] Matching the Schwarzschild region III to a DeSitter universe (region IV) at a surface $R=R_0=const.$ inside the horizon of region III.
    \item[(c)] Matching the Schwarzschild region III to an FRW universe (region IV) at a surface $T=T_B=const.$ inside the horizon of region III.
\end{itemize}
Notice that in all three cases the matching occurs inside the horizon for the region (III), and therefore it remains causally disconnected from the black hole exterior. We shall briefly discuss the cases (a) and (b) below and devote more attention to the case (c).

\subsection{Case a}

Here we consider the matching of two Schwarzschild regions (III given by the line element \eqref{sch} and IV with coordinates $\{\tilde{T},\tilde{R},\theta,\phi \}$ with the same line element and different value for the mass parameter). The matching is performed across the surfaces $R-R_0=0$ and $\tilde{R}-R_0=0$, with $R_0<2M_{\rm Sch}$ and $R_0<2\tilde{M}_{\rm Sch}$ respectively. Notice that having $(1-2M_{\rm Sch}/R_0)<0$ and $(1-2\tilde{M}_{\rm Sch}/R_0)<0$ implies that the matching is space-like in this case.
Continuity of the metric gives the relation
\be 
\frac{dT}{d\tilde{T}}=\sqrt{\frac{R(\tilde{R}-2\tilde{M}_{\rm Sch})}{\tilde{R}(R-2M_{\rm Sch})}},
\ee 
that holds on the boundary surface and can be easily integrated.
However, continuity of the extrinsic curvature can not be satisfied in this case, as we get
\bea 
 K_{TT}^i&=& \frac{M_{\rm Sch}}{R^2}\sqrt{1-\frac{2M_{\rm Sch}}{R}},\\
 K_{\theta\theta}^i&=&\frac{K_{\phi\phi}^i}{\sin^2\theta}= -R\sqrt{1-\frac{2M_{\rm Sch}}{R}},
\eea 
with $i=III, IV$.
Therefore one has to consider a non vanishing stress-energy tensor, which accounts for the difference in mass from one side to the other, that must be located at $R_0$.

\begin{figure}[t]
	\begin{center}
		\includegraphics[width=0.4\textwidth]{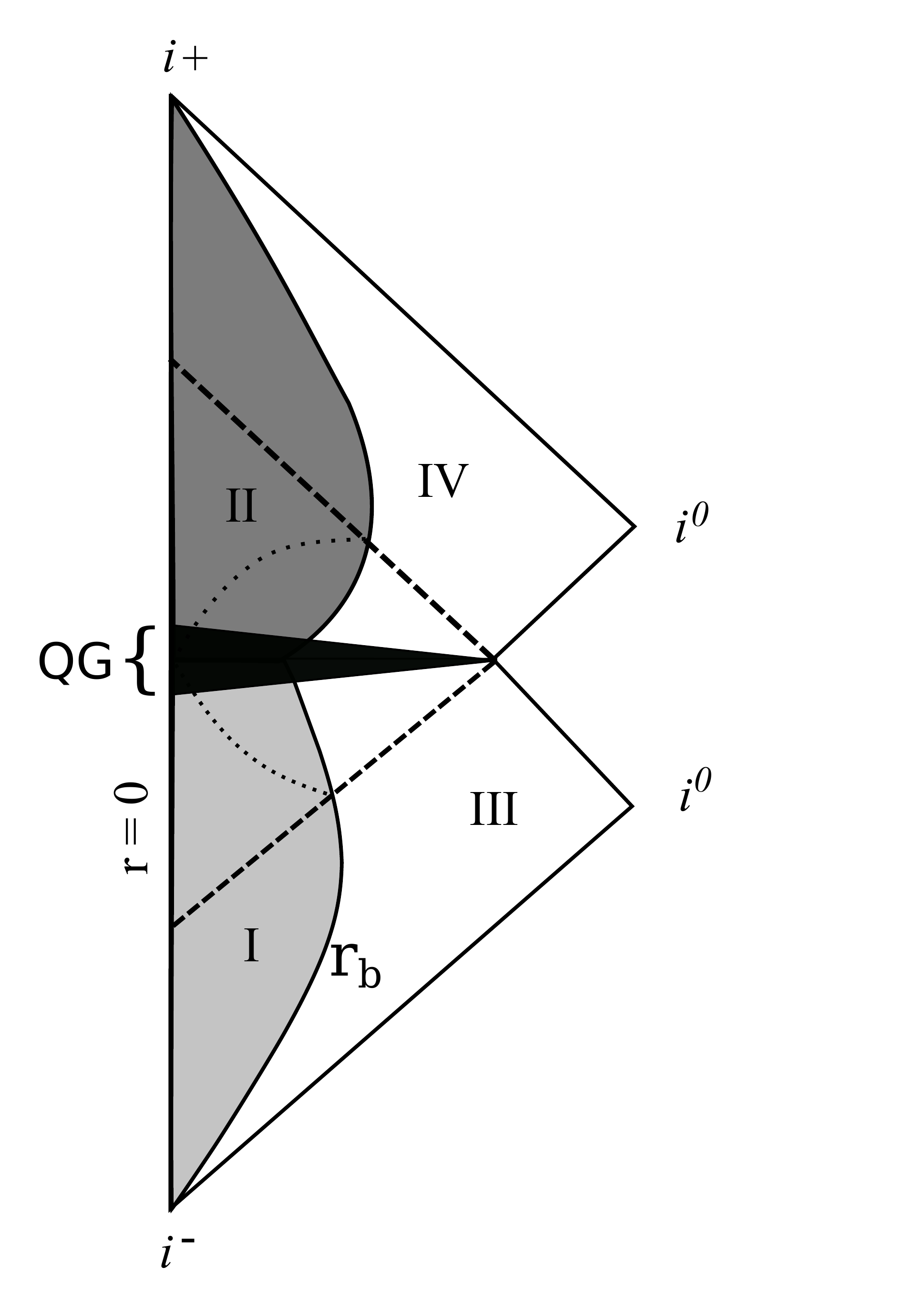}
	\end{center}
	\caption{The Penrose diagram for the described model in the case (a) contains four separate regions. Region I describes the collapsing OSD cloud matched to a vacuum Schwarzschild exterior, region III. Region II describes the expanding cloud which is matched to the collapsing one at the time of the bounce and to a different Schwarzschild exterior, region IV. The two Schwarzschild regions are causally disconnected as they are matched through a massive thin layer located at a constant radius inside the horizon. The dotted lines in region I and region II are apparent horizons. 
	\label{sch_p}}
\end{figure}

Matching region IV with region II can then be performed similarly to the matching of the region I and III. In this case, we obtain a dust cloud (i.e. $\omega=0$) which expands in a white-hole solution in a space-time causally disconnected from the parent universe. The Penrose diagram of this scenario is given in figure \ref{sch_p}.

\subsection{Case b}

\begin{figure}[t]
	\begin{center}
		\includegraphics[width=0.5\textwidth]{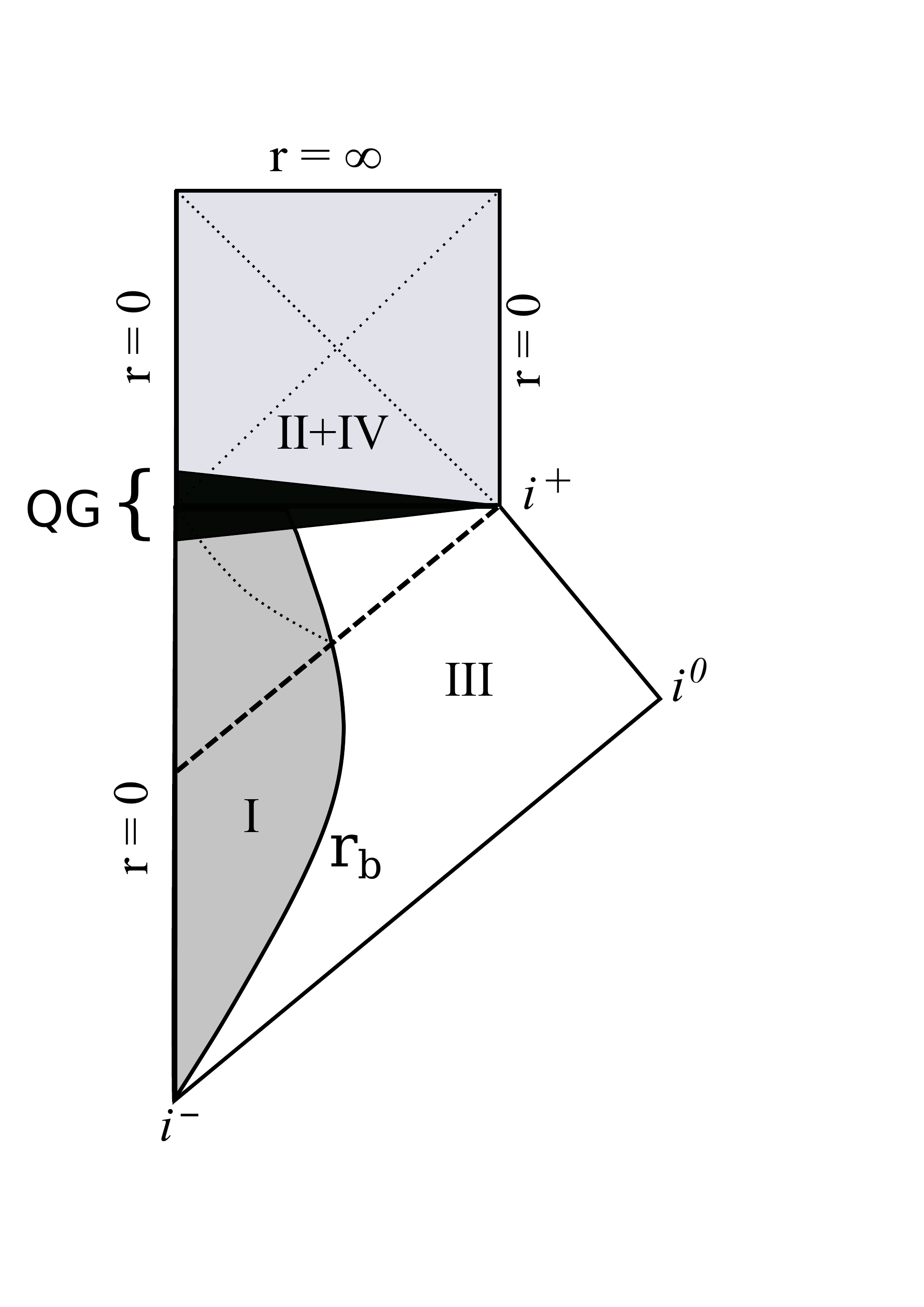}
	\end{center}
	\caption{The Penrose diagram for the model described in the case (b) shows distinct regions of space-time denoted as I, II+IV and III. Region I represents the modified OSD collapse, which is matched with an expanding FRW universe denoted by region II. The darker triangular region is the quantum gravity dominated region under which the collapsing and expanding space-times are joined. The exterior of region I is the vacuum Schwarzschild solution denoted by region III which is matched to a DeSitter universe denoted by region IV inside the horizon. In this model the expanding FRW universe and the DeSitter universe coincide. The dotted line in region I is the apparent horizon and the same in region II+IV are Cauchy horizons. \label{desit}}
\end{figure}

Here we consider the matching of the Schwarzschild region III to a DeSitter universe (region IV). The DeSitter line element can be written from the FRW metric \eqref{metric-in} in region II in the classical limit (i.e. $\rho_{\rm cr}\rightarrow +\infty$) when one takes $\omega=-1$. Therefore in this case region II and region IV coincide outside the QG dominated region and are both described by the FRW line element with $\omega=-1$.
The matching, neglecting the QG corrections, is again performed at a space-like surface given by $R-R_0=0$ inside the horizon on the Schwarzschild side and $t_{II}=t_B$ on the DeSitter side.
Continuity of the metric implies
\bea 
 \frac{dT}{dr_{II}}&=&\sqrt{\frac{r_{II}a_{II}(t_B)^3}{(1-k_{II}r_{II}^2)(2M_{\rm Sch}-r_{II}a_{II}(t_B))}},\\
R_0&=&r_{II}a_{II}(t_B).
\eea 
Once again the matching is in general not smooth as can be seen by the evaluation of the extrinsic curvature on the DeSitter side:
\bea 
 K_{rr}^{II}&=& \frac{a_{II}(t_B)\dot{a_{II}}(t_B)}{1-k_{II}r_{II}^2}, \\
 K_{\theta\theta}^{II}&=&\frac{K_{\phi\phi}^{II}}{\sin^2\theta}= a_{II}(t_B)\dot{a}_{II}(t_B)r_{II}^2,
\eea 
which vanishes at $t_{II}=t_B$ while the extrinsic curvature on the Schwarzschild side is non zero for $R_0\neq 2M_{\rm Sch}$.
This is the scenario that was considered in 
\cite{FMM1}
in the classical (i.e. without bounce) case.
The Penrose diagram of this scenario is given in \cite{FMM1} and, for the sake of comparison, we also report it herein, see figure~\ref{desit}. Notice that in \cite{FMM1} the radius where the transition occurred was determined by the limiting
curvature hypothesis, while in our case it is determined by the critical density.

\subsection{Case c}

Here we consider the matching between the Schwarzschild region III at a surface $T=T_B$ inside the horizon and the FRW universe described by region II at the surface $t_{II}=t_B$. 

The first thing to notice is that in this case, we are matching a time-like surface on one side to a space-like surface on the other. 
This is consistent with the idea that the light-cone structure of the geometry must change in the QG region near the bounce. 
{In the present case we will adapt an idea to construct the geometry interpolating from black hole to white hole that was proposed in~\cite{rov}.
In our case the transition is confined within the black hole horizon. We then construct an interpolating region (we shall call it region II.5) separating regions I and III from region II. 
First of all we need to make the assumption that the quantum-gravity dominated regime can still be described via a semi-classical geometry. Then we can assume that region II.5 is given by the generic spherically symmetric line element}
\begin{equation}
    ds^2_{II.5} = -A(\rti,\tti)d\tti^2 + B(\rti,\tti)d\rti^2 + C(\rti,\tti)d\Omega^2
\end{equation}
{with coordinates $ \{\tti,\rti,\theta,\phi\} $. The interpolating geometry shall have a finite thickness $\Delta\tilde{t}$. Then we match region III with region II.5 on a time-like hyper-surface $ \tti - T_B = 0 $ inside the horizon. By matching the first fundamental form we obtain,}
\begin{equation}
    \begin{aligned}
        & B(\rti,T_B) = \left( 1 - \frac{2M}{R} \right)^{-1}, \\
        & C(\rti,T_B) = R^2.
    \end{aligned}
\end{equation}
{The components of the second fundamental form in region III vanish, $ K_{RR} = K_{\theta\theta} = K_{\phi\phi} = 0 $, while the non-vanishing components of the second fundamental form as seen from the region II.5 are }
\begin{equation}
    \begin{aligned}
        & K_{\rti\rti} = -\frac{\dot{B}(\rti,T_B)}{2A(\rti,T_B)^{3/2}}, \\
        & K_{\theta\theta} = \frac{K_{\phi\phi}}{\sin^2\theta} =\frac{\dot{C}(\rti,T_B)}{2A(\rti,T_B)^{3/2}}.
    \end{aligned}
\end{equation}
{Therefore on the hyper-surface, we must impose $ \dot{B}(\rti,T_B) = \dot{C}(\rti,T_B) = 0 $. }

{Now we would like to match region II.5 to region II on a space-like hyper-surface characterized by the equation $ \tti - t_B = 0 $. The first junction conditions give}
\begin{equation}
    \begin{aligned}
        & B(\rti,t_B) = \frac{a(t_B)^2}{1-kr^2}, \\
        & C(\rti,t_B) = r^2 a(t_B)^2.
    \end{aligned}
\end{equation}
{The non vanishing components of the second fundamental form in region II are}
\begin{equation}
    \begin{aligned}
        & K_{rr} = \frac{a(t_B)\dot{a}(t_B)}{1-kr^2}, \\
        & K_{\theta\theta} = \frac{K_{\phi\phi}}{\sin^2\theta} = r^2 a(t_B)\dot{a}(t_B),
    \end{aligned}
\end{equation}
{while in region II.5 they are}
\begin{equation}
    \begin{aligned}
        & K_{\rti\rti} = \frac{\dot{B}(\rti,t_B)}{2A(\rti,t_B)^{3/2}}, \\
        & K_{\theta\theta} = \frac{K_{\phi\phi}}{\sin^2\theta} = \frac{\dot{C}(\rti,t_B)}{2A(\rti,t_B)^{3/2}}.
    \end{aligned}
\end{equation}
Therefore, since $ \dot{a}(t_B)=0 $, the second junction conditions become again $ \dot{B}(\rti,t_B) = \dot{C}(\rti,t_B) = 0 $.
%\begin{equation}
%    \begin{aligned}
%        & \frac{\dot{B}(\rti,\tti)}{2A(\rti,\tti)^{3/2}} = \frac{a(t_B)\dot{a}(t_B)}{1-kr^2} = 0, \\
%        & \frac{\dot{C}(\rti,\tti)}{2A(\rti,\tti)^{3/2}} = r^2 a(t_B)\dot{a}(t_B) = 0.
%    \end{aligned}
%\end{equation}
Notice that, as expected, the sign of $ B $ changes from one side to the other of the interpolating region. This describes the required change in the light-cone structure of the manifold in the QG region.
This is also the reason why the interpolating region II.5 can not be described by a quantum corrected FRW line element, satisfying the equation of motion \eqref{eom2}. In fact it is easy to check that the junction conditions for the quantum corrected FRW universe can be satisfied at the boundary between II and II.5 but not at the boundary between II.5 and III. This is in fact the reason why we need an interpolating geometry.
The thickness of the interpolating region, namely $\Delta\tilde{t}=T_B-t_B $ can then be considered to be arbitrarily small thus accomplishing the change of signature on the surface between region III and region II through a thin hyper-surface with signature change.

\begin{figure}[t]
	\begin{center}
		\includegraphics[width=0.5\textwidth]{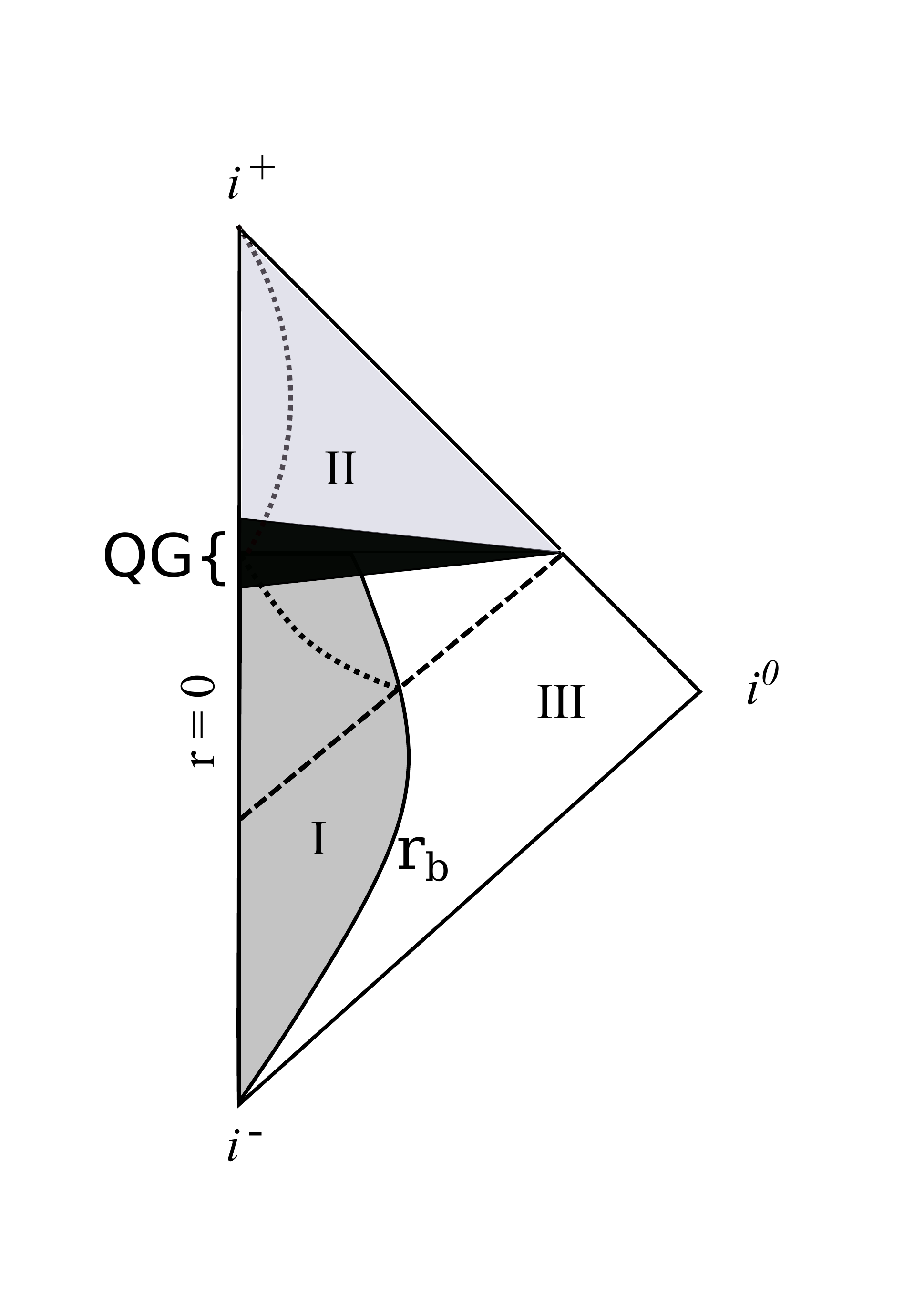}
	\end{center}
	\caption{The Penrose diagram for the model described in case (c) shows three distinct regions of space-time denoted as I, II and III. Region I represents the modified OSD collapse, which is matched with an expanding FRW universe denoted by region II. The darker triangular region is the quantum gravity dominated region (II.5) under which the collapsing and expanding space-times are joined. The exterior of region I is the vacuum Schwarzschild solution denoted by region III which is matched to interpolating region II.5 inside the horizon at a surface $T=T_B$, the quantum-gravity region II.5 is then matched to the expanding FRW region II at a surface $t_{II}=t_b$. The dotted lines in region I and region II represent the apparent horizons.
	\label{pen}}
\end{figure}

%In the present case, such a transition is confined to a thick shell separating regions I and III from region II. 
The matching of two manifolds through a hyper-surface with signature change was also studied in \cite{hellaby}.
%Continuity of the metric across the surface implies
%\bea
% \frac{dR}{dr_{II}}&=&\sqrt{\frac{a_{II}(t_B)^2}{1-k_{II}r_{II}^2}\left(1-\frac{2M_{\rm Sch}}{r_{II}a_{II}(t_B)}\right)}, \\
% R&=&r_{II}a_{II}(t_B),
%\eea 
%while continuity of the second fundamental form is ensured at the time of the bounce since all components of $K_{ab}$ on both sides vanish.
%
%However, as explained above, it is more realistic to picture the transition as occurring over the finite sized volume in which QG effects dominate. In our scenario, such a region is confined within the black hole horizon but extends outside the boundary of the collapsing cloud. Then the matching would occur across the hyper-surface of thickness $\Delta\tilde{t}$ that changes from space-like to time-like during the transition
\cite{mars}. The Penrose diagram representing our model is presented in Fig.\ref{pen}.      

We turn now the attention to the horizons in the collapsing and expanding phases. 
The apparent horizon for the collapsing (expanding) FRW metric, i.e. region I (II, respectively), is obtained from the expansion of future-directed outgoing and ingoing null geodesic congruences and it can be defined by the conditions
\begin{equation}\label{ahorcond}
    \begin{aligned}
        & \theta_l = 0, \\
        & \theta_n < 0.
    \end{aligned}
\end{equation}
Here,
\begin{equation}
    \begin{aligned}
        & \theta_l = \bigg[ g^{\mu\nu} + \frac{l^\mu n^\nu + n^\mu l^\nu}{-n^\sigma n^\lambda g_{\sigma\lambda}} \bigg] \nabla_\mu l_\nu,
    \end{aligned}
\end{equation}
where $ l_\mu $ and $ n_\mu $ are null tangents satisfying $ l^\mu l_\mu = n^\mu n_\mu = 0 $ and $ l^\mu n_\mu  = -2 $ and $ \theta_n $ is obtained from the above equation by replacing $ l_\nu $ with $ n_\nu $ in the covariant derivative.
In spherically symmetric space-times, the condition for the formation of trapped surfaces reduces to the requirement that the surface $ R(r,t)=const. $ is null, which is equivalent to requiring $ g^{\mu\nu}(\partial_{\mu}R)(\partial_{\nu}R)=0 $.

For the metric in equation \eqref{metric-in} the above condition becomes
\begin{equation}
    1-k_{i}r_{i}^{2}-\dot{R}_i^{2}=0.
\end{equation}
Notice that for $ k_{I}=0 $, the condition reduces to $r_{\rm ah}^i(t_i)=\pm 1/\dot{a}_i$ which is the usual equation for the apparent horizon in the homogeneous flat case. In general, the equation for apparent horizon radius becomes
\begin{equation}
    r_{ah}^i (t_i)= \frac{1}{\sqrt{k_{i}+\dot{a}_i^2}}.
\end{equation}
The occurrence of the bounce implies that $\dot{a}_i(t_B)=0$, so that the apparent horizon radius at the time of the bounce is
\be \label{ahb}
r_{\rm ah}^i(t_B)=\frac{1}{\sqrt{k_i}},
\ee 
which in the case of the collapsing phase is located outside the boundary of the cloud.

\begin{figure*}[t]
	\begin{center}
		\includegraphics[type=pdf,ext=.pdf,read=.pdf,width=5.5cm]{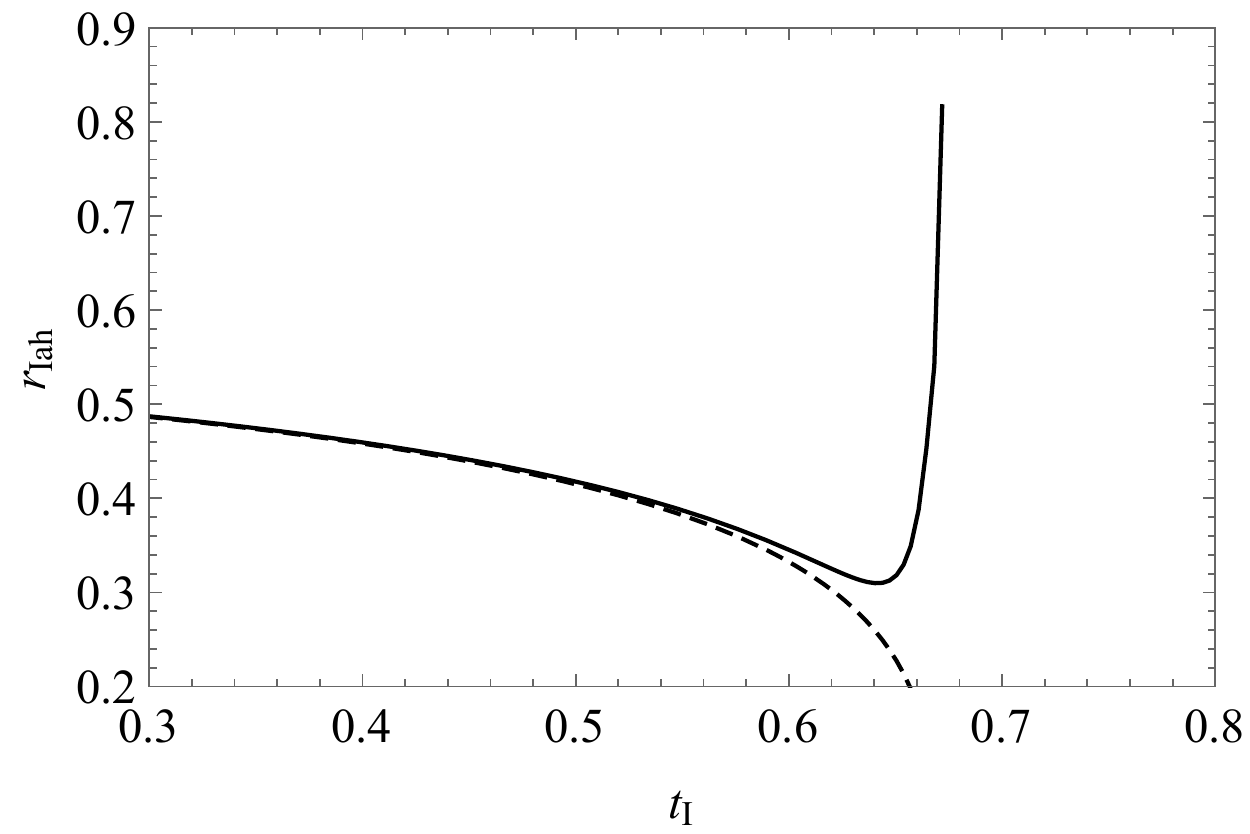}
		\includegraphics[type=pdf,ext=.pdf,read=.pdf,width=5.5cm]{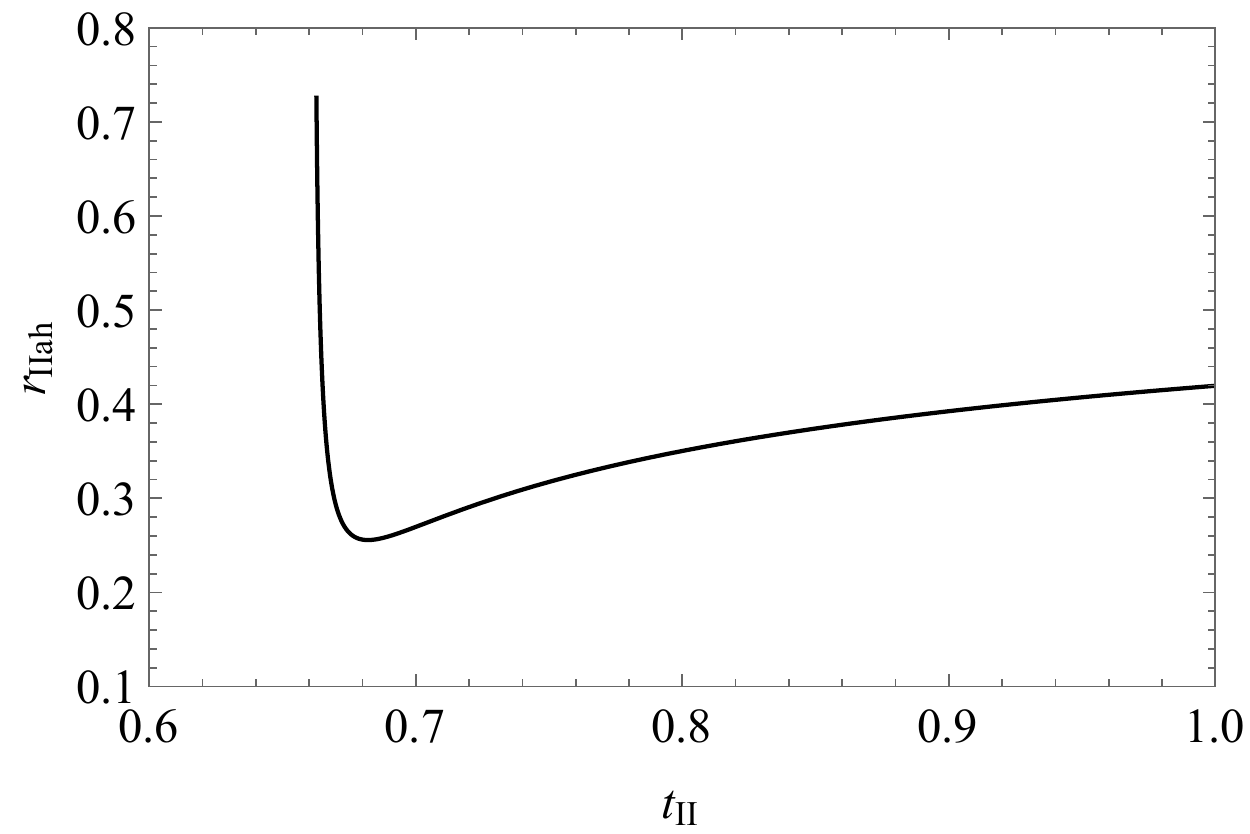}
		\includegraphics[type=pdf,ext=.pdf,read=.pdf,width=5.5cm]{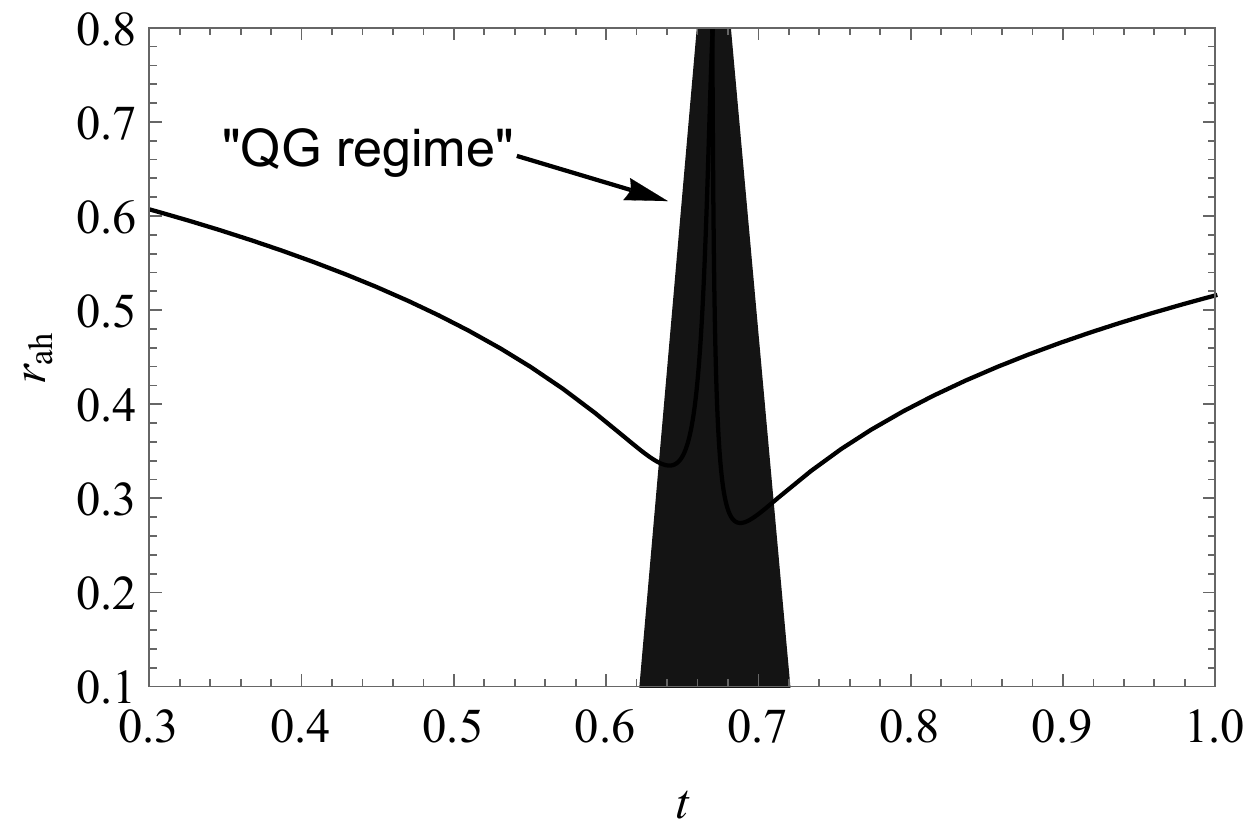}
	\end{center}
	\vspace{-0.5cm}
	\caption{ The left panel shows the apparent horizon curve $ r_{Iah}(t_{I}) $ for the collapsing phase with the solid line representing quantum inspired collapse and the dashed line representing classical collapse. The middle panel shows the apparent horizon curve $ r_{IIah}(t_{II}) $ for the expanding phase. In the right panel we plot a schematic of the apparent horizon curve both before and after collapse. The values of the parameters are $ k_I = 1.5 $, $ m_I = 2 $ and $ k_{II} = 1.96 $, $ m_{II} = 3 $, $ \omega = 0 $ for the collapsing and the expanding part respectively. \label{ahorcurve} }
\end{figure*}

Fig.~\ref{ahorcurve} shows the apparent horizon curves in the collapsing and expanding phases. The left panel in Fig.~\ref{ahorcurve} shows the apparent horizon curve $ r^I_{ah}(t_{I}) $ for the collapsing phase. The dashed line represents the apparent horizon curve for classical collapse without the bounce and the solid line represents the apparent horizon curve for quantum inspired collapse. We can see that in quantum inspired scenario, the apparent horizon reaches a constant value as given by equation (\ref{ahb}) at the time of bounce, which is not the case in classical collapse. The mid panel in Fig.~\ref{ahorcurve} shows the apparent horizon curve $ r^{II}_{ah}(t_{II}) $ for expanding universe after the bounce. The apparent horizon forms at a constant radius at the time of bounce and increases as the expansion proceeds. In the right panel, we show a schematic feature of apparent horizon formation combining both the curves in quantum inspired gravitational collapse.

\section{Discussion}\label{discuss}

Preliminary analysis of the data from the recent observation of the `shadow' of the supermassive black hole candidate at the center of M87 suggests that the object is well described by a black hole solution \cite{EHT1,EHT2}. However, there are reasons to believe that the singularities present in such solutions must be removed once the same object is described within a theory of Quantum-Gravity (see \cite{dm} and references therein).

In the present work, we suggested a possible reconciliation between the observation that astrophysical black holes are well described by the classical solutions and the theoretical implications that the removal of singularities within a theory of Quantum-Gravity may have for black holes. In particular, we constructed a dynamical model for an expanding universe forming inside a black hole by matching a collapsing spherically symmetric dust cloud with an expanding FRW universe on a space-like hyper-surface at the time when the collapse halts and the bounce occurs. 

Therefore, an observer at spatial infinity would only detect a classical black hole without any evidence of a bouncing event. For the expanding phase, we proposed three possibilities., namely, matching the exterior Schwarzschild region to another Schwarzschild space-time with a different value of the mass parameter, matching the exterior Schwarzschild region to a DeSitter universe and matching the exterior Schwarzschild to the expanding FRW universe inside the horizon. The second option has been explored in the literature before \cite{FMM1}. We showed the matching conditions in all three cases and calculated the apparent horizon for the collapsing matter and the expanding FRW universe.

\section*{Acknowledgement}
This work was supported by the Innovation Program of the Shanghai Municipal Education Commission, Grant No.~2019-01-07-00-07-E00035, the National Natural Science Foundation of China (NSFC), Grant No.~11973019, and Nazarbayev University Faculty Development Competitive Research Grant No.~090118FD5348. H.C. also acknowledges support from the China Scholarship Council (CSC), grant No.~2017GXZ019020. The work of A.A. is also supported in part by Projects No. VA-FA-F-2-008 of the Uzbekistan Ministry for Innovative
Development.  A.A. thanks Silesian University in Opava for the hospitality during his stay.

%-----------------------------------------------------

\end{document}